\documentclass[12pt]{article}
\usepackage{moretext}
\usepackage{graphicx}

\newtheorem{theorem}{Theorem}

\def\e{\mathop{\hbox{\sf E}}}
\def\expt#1{\e\left[#1\right]}

\def\exptB#1{\e\Bigl[#1\Bigr]}

\def\Prbak{{\rm Pr}}
\def\Pr#1{\Prbak\left\{#1\right\}}

\def\set#1{\left\{#1\right\}}
\def\setB#1{\Bigl\{#1\Bigr\}}

\def\proof{\par{\bf Proof}}
\newcommand{\select}{\chi}
\newcommand{\selection}{s}

\def\qed{%
    \relax\ifmmode
        \vrule height 8pt width 8pt depth 0pt
    \else
        {\unskip\nobreak\hfil\penalty50
        \hskip 2em\hbox{}\nobreak\hfil\vrule height 8pt width 8pt depth 0pt
        \parfillskip=0pt\par}
    \fi}

\def\myopenup#1pt{\advance\lineskip#1pt
  \advance\baselineskip#1pt
  \advance\lineskiplimit#1pt}
\def\myeqalign#1{\hbox{}\,\vbox{\myopenup7pt
  \everycr{}\tabskip=0pt
  \halign{\strut\hfil$\displaystyle{##}$%
	       &\hfil$\displaystyle{{}##{}}$\hfil
	       &     $\displaystyle{##}$\hfil
	\crcr#1\crcr}}\,}

\def\why#1#2{{\hbox to #1in{\hfill #2}}}
\def\eqref#1{(\ref{eq:#1})}
\def\eqlabel#1{\label{eq:#1}}

\begin{document}
\bibliographystyle{abbrv}
\bibstyle{clbiba}

\title{The Minimum Expectation Selection Problem}
\author{David Eppstein\thanks{Department of Information and Computer
Science, University of California, Irvine, Irvine, CA~~92697-3425,
{\tt eppstein@ics.uci.edu}.
Work supported in part by NSF 
grant CCR~99-12338.}\\
George S. Lueker\thanks{Department of Information and Computer
Science, University of California, Irvine, Irvine, CA~~92697-3425,
{\tt lueker@ics.uci.edu}.
Work supported in part by NSF grant CCR~91-19999 while visiting
DIMACS on a sabbatical in Fall 1998.}
}

\maketitle

\pagestyle{myheadings}
					
\def\implies{~~\Longrightarrow~~}

\newdimen\algindent \setbox0\hbox{{\bf then }} \algindent=0.5\wd0
\def\algorithm{\medskip\begingroup\leftskip=0in\parindent=0pt\parskip=1pt
\def\indent##1{\par\advance\leftskip by ##1\algindent\hangindent0.15in}
}
\def\endalgorithm{\par\medskip\endgroup}

\long\def\ignore#1{}

\def\atom{\hbox{\sl Atom}}

\begin{abstract}
We define the {\sl min-min expectation selection problem}
(resp.~{\sl max-min expectation selection problem})
to be that of selecting $k$ out of $n$ given discrete probability
distributions, to minimize (resp.~maximize) the expectation
of the minimum value resulting when independent 
random variables are drawn from the selected distributions.  
We assume each distribution has finitely many atoms.
Let $d$ be 
the number of distinct values in the support of the distributions.  We 
show that if $d$ is a constant greater than $2$, the min-min expectation 
problem is NP-complete but admits a fully polynomial time approximation 
scheme.  For $d$ an arbitrary integer, 
it is NP-hard to approximate the min-min 
expectation problem with any constant approximation factor.  The max-min 
expectation problem is polynomially solvable for constant $d$; we leave 
open its complexity for variable $d$.  We also show similar results for 
{\em binary selection problems} in which we must choose one distribution 
from each of $n$ pairs of distributions.
\end{abstract}

\section{Introduction}

Suppose we are given an integer $n$ and the distributions of $n$
independent random variables $Y_i$, for $1\le i\le n$.
We will let $l_0<l_1<\cdots<l_{d-1}$ denote the
possible values assumed by each of the random variables, and assume
these are specified in the input.  We assume that each
$Y_i$ is then specified in the input by giving, for each
$j\in{1,2,\ldots,d-1}$, the value of $\Pr{Y_i\ge l_j}$.
(Note that $\Pr{Y_i\ge l_0}$ need not be specified since it will always
be~1.)  For complexity purposes, we assume that all integer values in
the input are given in binary, and all other values are specified as
ratios of integers.  Let $N$ be the total number of bits required
to specify the input.

Suppose we wish to choose $k$ of the random variables so as to
minimize the expected value of the minimum of the selected variables.
More formally, if we wish to choose a subset
$S\subseteq\set{1,2,\ldots,n}$ with $|S|=k$ so as to minimize
\begin{equation} \eqlabel{minexpt}
\exptB{\min_{i\in S} Y_{i}},
\end{equation}
we call this the {\sl min-min expectation subset selection problem}.
We call the variation in which we wish to maximize \eqref{minexpt}
the {\sl max-min expectation subset selection problem}.

When we consider approximation results, we will assume
\begin{equation}
l_0\ge 0.				\eqlabel{lipositive}
\end{equation}
(Note that without this assumption we could use any approximation 
ratio bound to solve the decision problem, by translating all of the
$l_i$ appropriately.)

We also consider the problem of maximizing (resp.~minimizing) the
expected value of the maximum of the selected variables, which we call
the {\sl max-max} and {\sl min-max} problems.  By a simple change of
sign of all of the $l_i$, we see that the exact optimization forms of
the max-max and min-min problems, and of the min-max and max-min
problems, are equivalent in difficulty.  (Note that this equivalence
does not carry over to the approximation forms of the problems, since
the sign change would cause a violation of~\eqref{lipositive}.)

To motivate the min-min problem,
suppose you are a user of a peer-to-peer file sharing service such as 
Gnutella.  After performing a search, you have located several servers 
hosting copies of a file that you urgently need to download.  If bandwidth at 
your end of the network is not a limiting factor, you may be able to speed 
your download by requesting downloads from more than one server, and 
stopping when the first of these downloads reaches completion.  Suppose 
you have the capacity for $k$ simultaneous requests, and you have 
information from each server such as its connection type and echo time 
from which you can estimate a distribution on its download times.  Which 
$k$ servers do you choose in order to minimize the expected time until you 
have downloaded a complete copy of your file?

To motivate the min-max problem, consider an editor who wishes to select
$k$ referees out of $n$ qualified candidates so as to process an article
as quickly as possible.  Assume that the editor can estimate the time
used by each referee as a random variable, and assume that the times
used by each referee are independent.  Then, assuming the editor will wait
until all referee reports are received, s/he wishes to choose
a set of referees that minimizes the maximum of their times. 

Call the variations in which we are given $2n$ independent random
variables $Y_{i,\selection}$, for $1\le i\le n$ and $0 \le
\selection\le 1$, and asked to choose a function $\select:
\set{1,2,\ldots,n}\rightarrow\set{0,1}$ so as to optimize
\begin{equation}
\exptB{\min_{i=1}^{n} Y_{i,\select(i)}},
\end{equation}
the min-min expectation {\sl binary} selection problem or
the max-min expectation {\sl binary} selection problem.

We consider the complexity and approximability of these problems
both for the case in which $d$ is fixed and the case in which
$d$ is a parameter given in the input.
Our results hold for either the binary selection or subset selection
form of the problem.
In Section~\ref{sect:fixedd} we show that the variation of the min-min problem 
in which $d$
is fixed is NP-complete (assuming $d\ge3$), but admits a fully polynomial
time approximation scheme.  (See \cite{GareJohn} for definitions.)
Curiously, the max-min problem with fixed $d$
can be solved in polynomial time. 
In Section~\ref{sect:variabled} we show that when~$d$ is variable,
the min-min problem cannot be approximated to within any fixed ratio
in polynomial time unless $P=NP$;
we leave open the complexity of the max-min problem with variable~$d$.

The following formula, which follows from summation by parts, will be useful.
Suppose that a random variable $X$ assumes only the 
values $l_0,l_1,\ldots,l_k$.  Then 
\begin{equation} \eqlabel{partsexpt}
\expt{X} = l_0 + \sum_{j=1}^{d-1} (l_j - l_{j-1})\, \Pr{X\ge l_j} .
\end{equation}

\section{Fixed $d$} \label{sect:fixedd}
In this section we consider the case in which $d$ is fixed.  It will
be convenient to consider scaled negative logarithms of the probabilities: 
we describe a given random variable $X$ by a vector $\vec L=(L_1,L_2,
\ldots, L_{d-1})$ where $L_j = - \gamma^{-1} \ln \Pr{X\ge l_j},$ so
that
\begin{equation} \label{Lj.def}
\Pr{X\ge l_j} = e^{-\gamma L_j}.
\end{equation}
(Since $\Pr{X\ge l_0}$ will always be one, we need not specify the value
of~$L_0$.)
Here $\gamma$ is a positive value to be chosen later.
Note that the expectation of such a random variable is given by the
function 
\begin{equation} \eqlabel{fdef}
f(\vec L) = l_0 + \sum_{j=1}^{d-1} (l_j-l_{j-1}) e^{-\gamma L_j},
\end{equation}
and that this function~$f$ is convex, i.e., for $\alpha\in[0,1]$ and
arbitrary $\vec L$ and $\vec L'$, we have
$$f\bigl(\alpha \vec L + (1-\alpha)\vec L'\bigr)
	\le \alpha f(\vec L) + (1-\alpha) f(\vec L').$$
Note also that if $\vec L$ and $\vec L'$ specify the distributions
of two random variables, then the distribution of their minimum
is specified by $\vec L + \vec L'$.

Assume we are dealing with
the binary selection problem and
let $P_{i,\selection}$ be the vector of scaled negative logarithms 
corresponding to the random variable $Y_{i,\selection}$, i.e, 
the $j$th component $L_j$ of $P_{i,\selection}$
satisfies
$$\Pr{Y_{i,\selection} \ge l_j} = e^{-\gamma L_j}.$$
Then given a selection $\select: \set{0,1,\ldots,n-1}\rightarrow\set{0,1}$,
we have
\begin{equation} \eqlabel{exptasf}
\exptB{\min_{i=1}^{n} Y_{i,\select(i)}}
	= f\Bigl(\sum_{i=1}^{n} P_{i,\select(i)}\Bigr).
\end{equation}

\subsection{NP-completeness of the min-min problem}

The min-min problem is hard even when the all random variables must
assume values chosen from a set of size three.

\begin{theorem}
The min-min expectation binary selection 
problem is NP-complete for any fixed $d\ge 3$.
\end{theorem}

\proof.  Membership in NP is apparent.  To prove completeness
we perform a polynomial transformation from the subset-sum problem. 
Suppose we are given a set $\set{z_1,\ldots,z_n}$ of nonnegative
integers and asked whether the sum of some subset is equal to a given
integer~$T$.  Let $M$ be the maximum of the $z_i$; we will assume
that $T\le nM$ since otherwise the problem is trivial.
We show how to transform this problem into the min-min expectation
binary selection problem with $d=3$.  (The result for larger $d$
follows trivially by a padding argument.)

We choose $l_0=0$, $l_1=1$, and 
\begin{equation} \eqlabel{l2def}
l_2-l_1 = e^{2\gamma(nM-T)},
\end{equation}
where $\gamma$ is a positive number to be specified later.
For $1\le i\le n$ set
$$P_{i,0} = (0,2M)$$
and
$$P_{i,1} = (z_i,2M-z_i).$$
Note that each of these gives a valid set of probabilities; in particular,
since $0\le z_i\le M$ we have
$$1 \ge e^{-\gamma z_i} \ge e^{-\gamma (2M-z_i)} \ge 0.$$
We ask whether we can choose $\select$ so that the expected minimum
of the selected variables is at most $2e^{-\gamma T}$.

Now suppose that a selection function $\select$ has been specified
and let $S=\set{i~|~\select(i)=1}$.
Then using \eqref{fdef} and \eqref{exptasf},
and letting $\sigma=\sum_{i\in S}z_i$, we have
$$
\myeqalign{
\exptB{\min_{i=1}^{n} Y_{i,\select(i)}}
	&=& f\Bigl(\sum_{i=1}^{n} P_{i,\select(i)}\Bigr) 	\cr
	&=& l_0 + (l_1-l_0) \exp\Bigl(-\gamma\sum_{i\in S} z_i\Bigr)
	+ (l_2-l_1) \exp\left(-\gamma\Bigl(2nM-\sum_{i\in S} z_i\Bigr)\right) \cr
	&=& \exp\Bigl(-\gamma\sum_{i\in S}z_i\Bigr) + e^{2\gamma(nM-T)} 
            \exp\left(-\gamma\Bigl(2nM-\sum_{i\in S}z_i\Bigr)\right)	\cr
			&&\why4{(where we have used \eqref{l2def})}	\cr
	&=& e^{-\gamma\sigma} + e^{2\gamma(nM-T)} e^{-\gamma(2nM-\sigma)} \cr
	&=& e^{-\gamma\sigma} + e^{\gamma(\sigma-2T)}			\cr
	&=& e^{-\gamma T} \left(e^{\gamma(T-\sigma)} 
		+ e^{\gamma(\sigma - T)}\right).			\cr
}$$
Since $e^x+e^{-x}$ is minimized at $x=0$, it is clear that we can achieve
an expected minimum of $2e^{-\gamma T}$ if and only if we can choose
$S$ to make $\sigma = T$, i.e., if and only if the
answer to the subset sum problem is yes.

Two technical points need to be addressed.  First, since the
magnitudes of $z_i$, $T$, and $M$ can be exponentially large in the
length of the input, one might fear that this transformation would
produce an image of exponential length.  Second, of course, we cannot
output arbitrary reals in the constructed problem so we must use
finite precision and consider rounding problems.  To resolve these
problems, set $\gamma=1/(2nM)$; then it is easy to verify that no
number output is larger than $e+1$.  Also note that to resolve the
constructed decision problem it is sufficient to be able to
distinguish $e^{-\gamma T}(e^\gamma + e^{-\gamma})$ from $2e^{-\gamma
T}$.  It is easy to verify that we need only give a number of bits
that is polynomial in the input size to achieve this.\qed

This same construction shows that the min-min expectation subset selection
problem is also NP-complete: Since all $n$ of the variables
$Y_{i,0}$ constructed have the same distribution, picking any $n$
out of the $2n$ variables is equivalent to picking one from each pair.

Although the problem is NP-complete, the optimum solution can be easily 
approximated for fixed $d$, assuming that the $l_i$ are nonnegative.

\begin{theorem}
With nonnegative $l_i$ and a fixed value of $d$, the min-min expectation binary
selection problem admits a fully polynomial time approximation scheme.
\end{theorem}

\proof.  We use a standard rounding and dynamic programming approach.
Assume we are given some positive
\begin{equation} \eqlabel{epsbound}
\epsilon \le 1.
\end{equation}
Let the scale factor in (\ref{Lj.def}) be set at
\begin{equation} \eqlabel{gammadef1}
\gamma={\epsilon\over 6n}.
\end{equation}
Note that the components of each $P_{i,\selection}$ are positive
real numbers (or $+\infty$ when the corresponding probability is~0).
Let $\hat P_{i,\selection}$ be a vector in which each real component
$x$ of $P_{i,\selection}$ is rounded down to any integer in the range
$[x-2,x]$.  (The entries that are $\infty$ will not be rounded.  We
allow some flexibility in the rounding, rather than rounding to
$\lfloor x\rfloor$, to avoid having to do extremely precise
calculations when $x$ is very close to an integer.)  We will build a
table $A$ indexed by $d$-tuples of integers, specifying which vectors
are achievable as sums $\sum_{i=1}^{n} P_{i,\select(i)}$ for some
selection function $\select$.

A minor technical problem is that some components of $\sum_{i=1}^{n}
P_{i,\select(i)}$ may be infinite.  To deal with this we note that
the maximum value of any finite
component of $\sum_{i=1}^{n} \hat P_{i,\select(i)}$ is bounded by
$\gamma^{-1}\ln 1/p^*$, where $p^*$ is the product of all of the
nonzero probabilities given in the input.  Let $T$ be an integer between
$\gamma^{-1}\ln 1/p^*$ and $\gamma^{-1}\ln 1/p^*+2$.
Then we need not consider any finite indices of elements of $A$ which
exceed $T$.  Recalling that $N$ is the number of bits required to specify
the input, we see that $\ln 1/p^*$ is $O(N)$, so 
\begin{equation} \eqlabel{Tsize}
T = O(\gamma^{-1} N).
\end{equation}

Formally, we now define the value of
$A(s,L_1,L_2,\ldots,L_{d-1})$, for $0\le s\le n$ and 
$L_i\in\set{0,1,\ldots,T,T+1}$, to be a
boolean which is ${\bf true}$ if there exists a selection function
$\select$ such that for each $j$, the
$j$th component of $\sum_{i=1}^{s} \hat P_{i,\select(i)}$ is equal
to
$$\cases{
	L_j	&if $L_j\le T$, and	\cr
	\infty  &if $L_j=T+1$.		\cr
}$$
This table has $O(nT^{d-1})$ entries, and successive entries can
be computed in constant time by a standard dynamic programming approach.
Hence in view of \eqref{Tsize} and then \eqref{gammadef1}
the time required to build the table is 
$O(nT^{d-1}) = O\bigl(n(nN/\epsilon)^{d-1}\bigr)$, which is polynomial in
the input size and $\epsilon^{-1}$.

Once the table has been constructed, we simply evaluate the function
$f$ given in \eqref{fdef} at the tuples that the table tells us are
achievable.  More formally, let $g(\vec L)$ be a function which
maps components of $\vec L$ that are equal to $T+1$ back to
the infinity they represent, i.e., the $j$th component of
$g(L_1,L_2,\ldots,L_{d-1})$ is
$$\cases{
	L_j	&if $L_j\le T$, and		\cr
	\infty	&if $L_j = T+1$.		\cr
}$$
Then our estimate of the minimum expectation is
$$
E_{\rm min}= \min_{\vec L:~ A(n,\vec L)={\bf true}} f\bigl(g(\vec L)\bigr).
$$
Let $\vec L_{\rm min}$ be the value~of~$\vec L$ at which
the minimum is achieved, and let $\select_{\rm min}$ be a selection
that achieves this minimum, so
$$\sum_{i=1}^n\hat P_{i,\select_{\rm min}(i)} = \vec L_{\rm min}.$$
Since $\hat P$ was computed from $P$ by rounding down, and $f$ is a
decreasing function, we know that $E_{\rm min}$ is greater than or
equal to the true optimum.  Moreover, since each component of $\hat P$
was rounded down by at most~2, and we have considered all
possibilities for $\select$ when constructing $A$, from inspection
of $f$ we know that $E_{\rm min}$ exceeds the true minimum by a ratio
of at most
$$e^{2n\gamma} = e^{2n\epsilon/(6n)}
	= e^{\epsilon/3}
	\le 1+\epsilon/2,$$
where first step used \eqref{gammadef1}
and the last step used \eqref{epsbound}.

This achieves a ratio of $1+\epsilon/2$ assuming that the
computations are exact; it is not hard to verify
that the arithmetic can be done to only polynomially many places
and achieve a ratio of $1+\epsilon$.
\qed

Again, a similar algorithm gives an approximation scheme for the
min-min subset expectation selection problem.

\subsection{The max-min problem is in $P$}

We were surprised to find that, in contrast to the difficulty of the
min-min problem, the max-min expectation binary selection problem with
fixed $d$ can be solved in polynomial time.  

\begin{figure}
\includegraphics[width=6in]{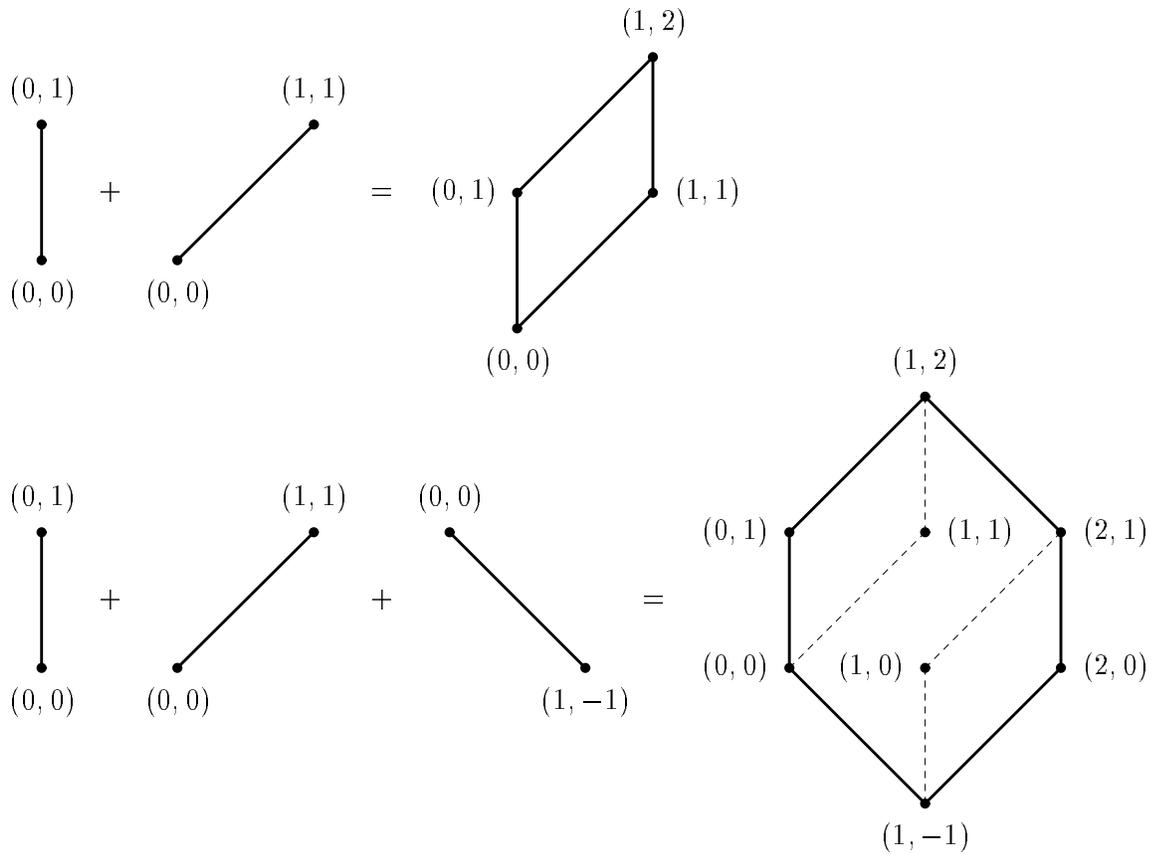}
\vskip 0.1in%
\caption[5]{\label{fig:1}Illustration of the construction of a 
zonotope.  We start with the interval from (0,0) to (0,1).
We sweep this by the second interval, from (0,0) to (1,1),
to form a rhombus.  We then sweep this rhombus by the third interval,
from $(0,0)$ to $(1,-1)$, to form a hexagon.  Note that $(1,0)$
and $(1,1)$ are not vertices of this hexagon.}
\end{figure}

Some background is useful.
Let $L_i$, for $1\le i\le n$, be line segments (considered as
point sets).
The {\sl Minkowski sum} of these line segments is
$$\setB{\sum_{i=1}^n v_i ~\Bigl|\Bigr.~ v_i\in L_i}.$$
The Minkowski sum of a set of line segments is called a {\sl zonotope}.
For example, see Figure~\ref{fig:1}, where, as suggested
by Edelsbrunner \cite{Edelsbrunner:bookAlgCombGeo}, the construction
of a zonotope is illustrated inductively.
If the $n$ line segments correspond to the unit vectors for each
of $n$ dimensions, the Minkowski sum is a hypercube in $n$ dimensions,
with $2^n$ vertices.
With fixed dimension $d-1$, the number of vertices in the Minkowski 
sum of $n$ segments grows much more slowly:
there are $O(n^{d-2})$ vertices, and they can be listed
in $O(n^{d-2}+n\log n)$ time \cite{Epp-MER-96}.

Going back to our problem, let
$$
S = \setB{ \sum_{i=1}^{n} \alpha_i P_{i,0} + (1-\alpha_i) P_{i,1}
        ~\Bigl|\Bigr.~ \alpha_i\in\set{0,1} }.
$$
We wish to find the maximum value of $f$ at any point in $S$.
Let $K$ be the convex hull of these points, so
$K$ is the zonotope
$$
K = \setB{ \sum_{i=1}^{n} \alpha_i P_{i,0}+(1-\alpha_i) P_{i,1}
        ~\Bigl|\Bigr.~
        \alpha_i\in[0,1] }.
$$
All vertices of $K$ are in $S$,
but not all points in $S$ are necessarily vertices of $K$.
Since $f$ is convex, we know that the maximum value of $f$ over
$S$ will be achieved at a vertex of $K$.  Thus we can find
the maximum in polynomial time by an exhaustive search
of the vertices of $K$.

Solving the max-min expectation subset selection problem (picking
$k$ out of $n$ variables) is done similarly, except that we need to
define 
$$
S = \setB{\sum_{i=1}^n \alpha_i P_i ~\Bigr|\Bigl.~
        \alpha_i \in \set{0,1} {\rm~~and~~}\sum_{i=1}^n \alpha_i = k}
$$
and
$$
K = \setB{\sum_{i=1}^n \alpha_i P_i ~\Bigr|\Bigl.~
        \alpha_i \in [0,1] {\rm~~and~~}\sum_{i=1}^n \alpha_i = k}.
$$
This $K$ is not in general a zonotope, but rather the intersection
of a zonotope with the hyperplane given by
$$\sum_{i=1}^n \alpha_i = k.$$
By results of \cite{BerEppGui-ESA-95}, we can still list the
vertices in polynomial time, and hence produce a polynomial time
algorithm by exhaustive search.

A technical point arises here: we have been assuming that we do
arithmetic with real numbers.  To show that the problem is in $P$, we
sketch here a proof that we need only carry the computations to $O(N)$
places of accuracy in order to determine the point at which the exact
optimum occurs.  First let $q$ be the product of all of the
denominators of fractions appearing in the input, and note that $\log
q = O(N)$.  It is then easy to see that the expectation of the minimum
of any subset of the random variables can be expressed as $i/q$ for
some integer $i$.  Thus to see where the maximum occurs, we only need
to be able to perform the computations accurately enough to be able to
correctly compare quantities that differ by at least $1/q$.  Since the
correct answer, even without the assumption that the $l_i$ are
positive, is bounded by $\max(|l_i|)$, it follows that
computations need only be done to $O(N)$ places of accuracy.

\section{Variable $d$} \label{sect:variabled}
\newcommand{\valuebase}{v}

\begin{theorem}
The problem of approximating the min-min expectation binary selection
with unrestricted $d$ to within any constant factor $r$ is NP-hard.
\end{theorem}

\proof.  We will transform CNF-sat to min-min expectation binary
selection in a way which gives a large gap between the optimum
solutions for satisfiable expressions and unsatisfiable expressions.

Suppose we are given a boolean formula $F$ on variables
$X_1,\ldots,X_n$, with clauses $C_0,C_1,\ldots,C_{c-1}$.  
We assume without loss of generality that $n\ge1$ and $c\ge1$.
For any literal $L$ and clause $C$, let $I(L,C)$ be 1 if $L$ is present in
$C$, and 0 otherwise.  Let $\atom(p,x)$ denote an atom of weight $p$
and location $x$.  Choose
\begin{equation}\eqlabel{pdef}
p = {1\over r(c+1)}
\end{equation}
and
\begin{equation}\eqlabel{ddef}
\valuebase={r^n(c+1)}^n=p^{-n}.
\end{equation}
Construct a min-min expectation binary selection problem $M$ with, for
$1\le i\le n$, variables
\begin{equation}\eqlabel{y1}
Y_{i,1} \hbox{ with distribution } \atom(p^c,\valuebase^c) 
	+ \sum_{\selection=0}^{c-1} \atom\left((1-p)p^\selection,\valuebase^{\selection-I(X_i,C_\selection)}\right)
\end{equation}
and
\begin{equation}\eqlabel{y0}
Y_{i,0}  \hbox{ with distribution }  \atom(p^c,\valuebase^c)
    + \sum_{\selection=0}^{c-1} \atom\left((1-p)p^\selection,\valuebase^{\selection-I(\overline X_i,C_\selection)}\right).
\end{equation}
We show that if $F$ is satisfiable then the optimum solution for~$M$
is at most $1/r$, but if $F$ is unsatisfiable the optimum solution
is at least 1.

To begin, fix integers $i$ and $\selection$, with $0\le \selection<c$,
and consider the probability that a single variable $Y_{i1}$ is at
least $\valuebase^\selection$.  If the literal $X_i$ is not in clause
$C_\selection$, then from \eqref{y1} the total weight of all atoms in
$Y_{i1}$ with locations at or beyond $\valuebase^\selection$ is $p^c +
\sum_{l=\selection}^{c-1} (1-p)p^l = p^\selection$.  If, however, the
literal $X_i$ is in clause $C_\selection$, then a term
$(1-p)p^\selection$ disappears from this sum so $Y_{i1}$ is at least
$\valuebase^\selection$ with probability only $p^{\selection+1}$.  In
summary,
\begin{equation}\eqlabel{y12}
\Pr{Y_{i1}\ge \valuebase^\selection} = \cases{
		p^{\selection+1} &if $X_i$ is in $C_\selection$		\cr
		p^\selection     &if $X_i$ is not in $C_\selection$.	\cr
		}
\end{equation}
Similarly from \eqref{y0}
\begin{equation}\eqlabel{y02}
\Pr{Y_{i0}\ge \valuebase^\selection} = \cases{
	p^{\selection+1}&if $\overline X_i$ is in $C_\selection$	\cr
	p^\selection	&if $\overline X_i$ is not in $C_\selection$.	\cr
	}
\end{equation}

Next consider the probability that the minimum of the selected
variables is at least $\valuebase^\selection$, i.e.,
$$
\Pr{\min_{i=1}^n Y_{i,\select(i)} \ge \valuebase^\selection}
= \prod_{i=1}^n \Pr{Y_{i,\select(i)} \ge \valuebase^\selection}.
$$
Using \eqref{y12} and \eqref{y02}, if the assignment $X_i=\select(i)$
does not satisfy clause $C_\selection$, this product is just $p^{sn}$.
If, on the other hand, the assignment does satisfy clause
$C_\selection$, then at least one of the terms in the product will be
bounded by $p^{\selection+1}$, so the product is at most $p^{sn+1}$.
Summarizing,
\begin{equation}\eqlabel{ydj}
\Pr{\min_{i=1}^n Y_{i,\select(i)} \ge \valuebase^\selection} {\rm~~~is~~~} 
\cases{
	{}=p^{sn}&	if $\select$ does not satisfy $C_\selection$\cr
	{}\le p^{sn+1}&	if $\select$ satisfies $C_\selection$.\cr
	}
\end{equation}

Finally, consider the expectation of the minimum of the selected
variables.  Note that the possible values for the random variables are
$v^{-1},1,v,v^2,\ldots,v^{c-1}$, so using \eqref{partsexpt} we have
$$
\expt{\min_{i=1}^n Y_{i,\select(i)}}  
	= \valuebase^{-1} 
		+ \sum_{\selection=0}^{c-1} \Pr{\min_{i=1}^n Y_{i,\select(i)}
	\ge \valuebase^\selection}
		(\valuebase^\selection-\valuebase^{\selection-1}).
$$
Using \eqref{ydj}, we see that if $\select$ satisfies all of the
clauses, this becomes
\begin{equation} \eqlabel{upbound}
\myeqalign{
\expt{\min_{i=1}^n Y_{i,\select(i)}}
	&\le& \valuebase^{-1} 
	  + \sum_{\selection=0}^{c-1}p^{sn+1}(\valuebase^\selection-\valuebase^{\selection-1})		\cr
	&=&   \valuebase^{-1} + (1-\valuebase^{-1}) \sum_{\selection=0}^{c-1}p^{sn+1}\valuebase^\selection		\cr
	&=&   p^n + (1-p^n) \sum_{\selection=0}^{c-1}p				\cr
	&&\hskip1in\hbox{(by using \eqref{ddef})}			\cr
	&=&   p^n + (1-p^n)cp.
}
\end{equation}
Since $n\ge1$ this is less than
$$p+cp = p (c+1) = {1\over r}.
$$ 
If, on the other hand, $\select$ violates some clause,
say $C_\selection$, again using \eqref{ydj} we have
\begin{equation} \eqlabel{lobound}
\myeqalign{
\expt{\min_{i=1}^n Y_{i,\select(i)}}
	&\ge& \valuebase^{-1} 
	  + p^{sn}(\valuebase^\selection-\valuebase^{\selection-1})	\cr
	&=& \valuebase^{-1} 
	  + p^{sn}(1-\valuebase^{-1})\valuebase^\selection		\cr
	&=& p^n + (1-p^n) = 1.						\cr
	&&\hskip1in\hbox{(by using \eqref{ddef})}			\cr
}
\end{equation}
Thus approximating the solution to within a factor of $r$ would enable
us to distinguish between satisfiable and unsatisfiable expressions.\qed

\begin{theorem}
The problem of approximating the min-min expectation subset selection
with unrestricted $d$ to within any constant factor $r$ is NP-hard.
\end{theorem}

\proof~sketch.  The
proof is similar to that of the previous theorem, so we only briefly
describe the necessary changes.  We set
\begin{equation}\eqlabel{pdef2}
p = {1\over r(n+c+1)}.
\end{equation}
Now, for $1\le i\le n$, we construct variables $Y_{i,1}$
with distribution
\begin{equation}\eqlabel{y1b}
\atom(p^{c+n},\valuebase^{c+n}) 
    + \sum_{\selection=0}^{c-1} \atom\left((1-p)p^\selection,\valuebase^{\selection-I(X_i,C_\selection)}\right)
    + \sum_{\selection=0}^{n-1} \atom\left((1-p)p^{c+\selection},\valuebase^{c+\selection-\delta_{i,\selection}}\right)
\end{equation}
and variables $Y_{i,0}$ with distribution
\begin{equation}\eqlabel{y0b}
\atom(p^{c+n},\valuebase^{c+n})
    + \sum_{\selection=0}^{c-1} \atom\left((1-p)p^\selection,\valuebase^{\selection-I(\overline X_i,C_\selection)}\right)
    + \sum_{\selection=0}^{n-1} \atom\left((1-p)p^{c+\selection},\valuebase^{c+\selection-\delta_{i,\selection}}\right),
\end{equation}
where as usual we define
$$
\delta_{i,\selection} = \cases{
		1&if $i=\selection$\cr
		0&otherwise.\cr
		}
$$
The problem we construct is to choose $n$ of these $2n$ variables so
as to minimize the expected value of the minimum of the selected
variables; let $Z$ be distributed as the minimum of the
selected variables.  If for some $i$ neither of $Y_{i0}$ and $Y_{i1}$ is
selected, arguing much as before we have
$\expt{Z} \ge 1.$
Thus in order to get an expectation below~1 we must pick at least one
of each pair.  But since there are $n$ pairs and we must pick exactly
$n$ variables, this means we can also pick at most one, and hence
must pick exactly one, from each pair.
Again as before, if the selected variables do not correspond to a satisfying
assignment, we will have $\expt{Z} \ge 1.$
On the other hand, if they do correspond to a satisfying assignment, we
will have 
$$\expt{Z} \le p^n + (1-p^n)(c+n)p < p + (c+n)p = {1\over r}.$$
Thus approximating the solution to within a factor of $r$ would again enable
us to distinguish between satisfiable and unsatisfiable expressions.\qed

\section{Acknowledgements}
The second author wishes to thank Mike Fredman and Larry Larmore
for a discussion which ultimately lead to his interest in this problem.

\bibliography{/home/lueker/biblio/all}

\end{document}